\documentclass[twocolumn,twoside,prd,floatfix,letterpaper]{revtex4}

\usepackage{ifpdf}
\ifpdf 
  \usepackage[pdftex]{graphicx} 
\else 
  \usepackage{graphicx} 
\fi
\usepackage{amsmath}
\usepackage{amssymb}
\usepackage{fancyhdr}
\usepackage{journals}
\usepackage{rotating}
\usepackage[hyperindex]{hyperref} 

\def \Bard {{\sc Bard}}

\def \Quaero {{\sc Quaero}}

\begin{document}

\title{A Quantitative Measure of Experimental Scientific Merit}
\author{Bruce Knuteson}
\homepage{http://mit.fnal.gov/~knuteson/}
\email{knuteson@mit.edu}
\affiliation{MIT}


\begin{abstract}
Experimental program review in our field may benefit from a more quantitative framework within which to quantitatively discuss the scientific merit of a proposed program of research, and to assess the scientific merit of a particular experimental result.  This article proposes explicitly such a quantitative framework.  Examples of the use of this framework in assessing the scientific merit of particular avenues of research at the energy frontier in many cases provide results in stark contradiction to accepted wisdom.  The experimental scientific figure of merit proposed here has the potential for informing future choices of research direction in our field, and in other subfields of the physical sciences.
\end{abstract}

\maketitle
\tableofcontents

\section{Motivation}

In the context of determining which research programs to pursue, review committees often must decide the relative scientific merits of proposed experiments.  Within large experiments, deciding which analyses to emphasize requires similar decisions.  These issues arise in the discussion of experiments or analyses in which the result is not yet known.

A related issue is faced by those assessing the scientific merit of an experiment or analysis whose result is known.  This topic is the subject of much innocuous lunchroom conversation, and more seriously in the evaluation of the organizations and individuals responsible for producing the result.

In such a quantitative field as experimental high energy physics, the discussions leading to these decisions and evaluations are notably non-quantitative.

The intent of Sec.~\ref{sec:InformationEntropy} is to suggest a quantitative framework for such discussions, including a specific figure of merit to quantify the scientific merit of a proposed experimental research program, and a specific figure of merit to quantify the scientific merit of a completed experimental result.  Sec.~\ref{sec:Examples} provides explicit worked examples.

\section{Information content}
\label{sec:InformationEntropy}

Given the academic and esoteric nature of our field, the appropriate figure of merit is not expected economic payoff, but rather should be formulated in terms of an experiment's information content.  Given the rather grandiose nature of our field, we will not concern ourselves with distinguishing qualitatively equivalent but quantitatively different states of understanding.  Whether a Standard Model Higgs boson exists or not are considered two qualitatively different states of understanding; if a Standard Model Higgs boson exists, whether it has mass 115~GeV or 160~GeV are considered two qualitatively equivalent states of understanding.

The appropriate figure of merit to measure an experiment's information content is well known in the context of information theory.  The original reference is Shannon's classic 1948 paper, ``A Mathematical Theory of Communication''~\cite{Shannon}. 
 The essential idea behind this figure of merit is developed in Sec.~\ref{sec:InformationEntropy:Basic}, with a more sophisticated version developed in Sec.~\ref{sec:InformationEntropy:Advanced}.

\subsection{Basic formulation}
\label{sec:InformationEntropy:Basic}

The value of a particular experimental result in an academic field clearly should be measured by how much is learned from the result.  Equivalently, the value of a result is how surprised you are that the particular result has been obtained.

Suppose an experiment can have one of two outcomes.  Denote the two possibilities by $X=\{x_1,x_2\}$, and let $p(x_1)$ and $p(x_2)$ represent the expectation of the two possible outcomes $x_1$ and $x_2$ being realized, with $p(x_1)+p(x_2)=1$.  For example, assigning $p(x_1)=0.8$ and $p(x_2)=0.2$ represents betting odds of 4:1 in favor of outcome $x_1$ being obtained.  In general, an experiment will have $n$ possible outcomes $X={x_1,\ldots,x_i,\ldots,x_n}$, with each outcome $x_i$ having expectation $p(x_i)$.

The surprise of obtaining the particular result $x_i$ is clearly large if $p(x_i)$ is small.  The scientific value of the result $x_i$ is therefore large if $p(x_i)$ is small.  The scientific value of the result $x_i$ is small if $p(x_i)$ is near unity, since in this case little has been learned.  The scientific value of obtaining the result $x_i$ must thus be a monotonically decreasing function of $p(x_i)$.

A good measure of scientific merit should have the desired property of addition:  the scientific merit of a paper announcing the discovery of a Standard Model Higgs boson plus the scientific merit of a paper announcing the discovery of a new $Z'$ boson should equal the scientific merit of a paper containing both discoveries.  Equivalently, the act of stapling two papers together should not change their total scientific merit.  

The scientific merit of a particular result $x_i$ should thus (i) be a monotonically decreasing function of the expectation $p(x_i)$ that the result would be obtained, and (ii) be appropriately additive.  This leads uniquely to the choice~\cite{Shannon}
\begin{equation}
\label{eqn:eqn1}
\text{surprisal}(x_i)= -\log_{10}{p(x_i)}
\end{equation}
for the scientific merit of obtaining the result $x_i$.  The negative sign in Eq.~\ref{eqn:eqn1} is required for the scientific merit to be a monotonically decreasing function of $p(x_i)$, and the logarithm is required for the scientific merit to have the desired behavior under addition.  The use of logarithm base ten is arbitrary but convenient, and corresponds to a choice of units~\footnote{Choice of logarithm base corresponds to a choice of units.  In communication theory the base 2 logarithm is commonly used, allowing interpretation in units of bits.  For the derivation of analytic results the natural logarithm is convenient.  In this article the decimal logarithm is used for its intuitive interpretation.}~\footnote{As a sanity check, consider an experiment with ten possible outcomes $X=\{x_1,\ldots,x_i,\ldots,x_{10}\}$, each of which has equal expectation, so that $p(x_i)=0.1$ for all $i\in\{1,\ldots,10\}$.  Obtaining the result $x_7$ corresponds to a surprisal of $\text{surprisal}(x_7)=-\log_{10}{0.1}=1$.  Consider a second experiment with one hundred possible outcomes $X=\{x_1,\ldots,x_i,\ldots,x_{100}\}$, each of which has equal expectation, so that $p(x_i)=0.01$ for all $i\in\{1,\ldots,100\}$.  Obtaining the result $x_{77}$ corresponds to a surprisal of $\text{surprisal}(x_{77})=-\log_{10}{0.01}=2$.  Consider a third experiment corresponding to a collaboration of the first and second experiments with one thousand possible outcomes $X=\{(x_1,x_1),\ldots,(x_1,x_{100}),(x_2,x_1),\ldots,(x_{10},x_{100})\}$.  Obtaining the joint result $(x_7,x_{77})$ corresponds to a surprisal of $\text{surprisal}(x_7,x_{77})=-\log_{10}{0.001}=3=1+2$, as desired.}.

This article suggests that the surprisal of a particular experimental result is the appropriate measure of scientific merit for an experiment or analysis whose result is known.  The larger the surprisal of a particular outcome, the greater its ({\em{a posteriori}}) scientific merit.  

A measure of the scientific merit of an experimental result has thus been obtained in Eq.~\ref{eqn:eqn1}.  The definition of rational behavior is decision making that maximizes expected payoff.  The payoff of obtaining any particular result $x_i$ is given by Eq.~\ref{eqn:eqn1}.  The expected payoff is thus the payoff of each possible result $x_i$, weighted according to the expectation that the result $x_i$ will be obtained.  The expected payoff of an experiment with possible outcomes $X={x_1,\ldots,x_i,\ldots,x_n}$ is thus
\begin{equation}
\label{eqn:eqn2}
H(X)= \sum_i{p(x_i) \, \text{surprisal}(x_i)} = - \sum_i{p(x_i) \log_{10}{p(x_i)}}. 
\end{equation}
The expected payoff $H(X)$ is thus the appropriate scientific merit of a proposed experiment, before the result is known.

Nearly all of the discussion so far can be found in the first few pages of any introductory information theory text.  The only new idea in this article is that the surprisal and expected surprisal of Eqs.~\ref{eqn:eqn1} and~\ref{eqn:eqn2} can actually be applied in practice to quantify the scientific merit of completed and proposed high energy physics experiments.

\subsection{Advanced formulation}
\label{sec:InformationEntropy:Advanced}

The previous section introduced the essential idea of this article.  This section completes the development of necessary concepts.

In the previous section it was implicitly assumed that each possible experimental outcome corresponds one-to-one with a qualitatively new state of knowledge.  This is clearly not always the case:  flipping a coin can result in heads or tails, and thus obtaining heads has a nonzero surprisal according to Eq.~\ref{eqn:eqn1}, but obtaining heads does not reveal much about how Nature works.

It is thus useful to consider a set of qualitatively distinct and mutually incompatible states of knowledge $Y=\{y_1,\ldots,y_j,\ldots,y_m\}$, the $j^{\text{th}}$ of which is generally accepted to be correct with probability $p(y_j)$, in the usual Bayesian interpretation.  Since the process of normal science relies to a significant degree on a scientific community's shared view of important problems and possible solutions, in practice there is little difficulty defining $Y$~\cite{Kuhn}.  The possible states of knowledge are complete and mutually exclusive, so that $\sum_{j=1}^{m}{p(y_j)}=1$~\footnote{The generalization to a continuous set of possibilities is straightforward.}.  An example might be $Y=\{y_1,y_2\}$, where $y_1$ corresponds to the state of knowledge in which a Higgs boson is known to exist, and $y_2$ corresponds to a state of knowledge in which a Higgs boson is known not to exist.  An experiment may have possible outcomes $X=\{x_1,x_2\}$, where $x_1$ corresponds to the result that a Higgs boson is found, and $x_2$ corresponds to the result that no Higgs boson is found.  Depending on the sensitivity of the experiment, $x_1$ and $x_2$ may not be in one-to-one correspondence with $y_1$ and $y_2$.  A null result from a Higgs boson search at the Tevatron, which is limited by the amount of data collected and the collider energy, will not prove the Higgs boson does not exist.

A slightly more sophisticated understanding of Sec.~\ref{sec:InformationEntropy:Basic} is required.  Noting that what counts is the state of knowledge rather than the particular experimental outcome, Eq.~\ref{eqn:eqn2} should be written in terms of $Y$ rather than $X$:
\begin{equation}
H(Y) = - \sum_{j}{p(y_j) \log_{10}{p(y_j)}}.
\label{eqn:informationEntropy_prior}
\end{equation}
The quantity $H(Y)$ can be considered to be the entropy associated with the state of knowledge $Y$.  The state of knowledge $Y$ is a superposition of orthogonal microstates $y_j$, each with expectation $p(y_j)$.  It was implicitly assumed in Sec.~\ref{sec:InformationEntropy:Basic} that the experiment under consideration results in a definitive state of knowledge, where one microstate $y_j$ is selected, corresponding to zero information entropy after the result is obtained.  In general, an experiment may result in a state of knowledge with nonzero information entropy: if the Higgs boson is not found at the Tevatron, the resulting state of knowledge will still consist of a superposition of the two possibilities that the Higgs boson does or does not exist, with slightly adjusted relative expectations.

Assume the experimental result $x_i$ is obtained.  The expectation that $y_j$ is correct was $p(y_j)$ before the experiment, and is $p(y_j|x_i)$ after the experiment~\footnote{Standard Bayesian notation is used.  The joint probability $p(x_i,y_j)$ is a number between zero and unity quantifying the degree of belief that the outcome of the experiment will be $x_i$ and that the true hypothesis is $y_j$.  The conditional probability $p(y_j|x_i)$ is a number between zero and unity quantifying the degree of belief that the true hypothesis is $y_j$, given that the outcome of the experiment is $x_i$.}.  The information entropy after the experimental result $x_i$ is obtained will be
\begin{equation}
H(Y|x_i) = - \sum_{j}{p(y_j|x_i) \log_{10}{p(y_j|x_i)}},
\label{eqn:informationEntropy_posterior_xi}
\end{equation}
in analogy with Eq.~\ref{eqn:informationEntropy_prior}.

Before the experiment is performed, it is not known what result will be obtained.  Before the experiment is performed, the expected information entropy that will remain after the experiment is performed is
\begin{eqnarray}
H(Y|X) & = & \sum_{i}{p(x_i) \, H(Y|x_i)} \nonumber \\
       & = & - \sum_i {p(x_i) \sum_{j}{p(y_j|x_i) \log_{10}{p(y_j|x_i)}}} \nonumber \\
       & = & - \sum_{i,j} {p(x_i,y_j) \log_{10}{p(y_j|x_i)}}.
\label{eqn:informationEntropy_posterior}
\end{eqnarray}
Performing an experiment $X$ is thus expected to reduce the information entropy of our state of knowledge $Y$ from $H(Y)$ to the conditional entropy $H(Y|X)$.

Before the experiment is performed, the information entropy $H(Y)$ quantifies our degree of uncertainty.  Performing the experiment $X$ is expected to reduce this uncertainty to a smaller information entropy $H(Y|X)$.  The information expected to be gained by performing the experiment $X$ is thus
\begin{equation}
\Delta H=H(Y)-H(Y|X).
\label{eqn:deltaInformationEntropy}
\end{equation}

This article suggests that the expected information content in a particular experiment or analysis, measured in terms of the expected reduction in information entropy $\Delta H$, is the appropriate measure of scientific merit for an experiment or analysis whose result is not yet known.  The larger the expected information content of an experiment, the greater its ({\em{a priori}}) scientific merit.

Experiments that are guaranteed to result in a state of knowledge of zero entropy have $H(Y|X)=0$.  For such experiments, the information content $\Delta H$ is equal to the entropy $H(Y)$ of the initial state of knowledge.

Assume the state of knowledge $y_j$ is actually correct.  The scientific merit of obtaining a particular result $x_i$ should then be the additional evidence 
\begin{equation}
\text{evidence}(y_j|x_i) = \log_{10}{\frac{p(y_j|x_i)}{p(y_j)}}
\label{eqn:evidence}
\end{equation}
that the experiment can provide in favor of the (correct) state of knowledge $y_j$.  The logarithm again ensures the desired behavior under addition.  As an example, $p(y_j)=0.09$ and $p(y_j|x_i)=0.90$ results in $\text{evidence}(y_j|x_i) = 1$, and the experimental result $x_i$ corresponds to one unit of evidence in favor of $y_j$.

Since in practice the state of knowledge $y_j$ is not known to be correct, the scientific merit of obtaining the experimental result $x_i$ is the evidence the result $x_i$ provides in favor of $y_j$, averaged over possibly correct states of knowledge $y_j$, weighted according to the new expectation that each $y_j$ is correct:~\footnote{Terms in the sum over $j$ for which $p(y_j|x_i)=0$ vanish.}~\footnote{Since high energy physicists may find the usual notations for the Kullback-Leibler divergence, such as $D_{KL}( p(Y|x_i) || p(Y) )$, to be somewhat cryptic, we have opted to spell out ``information gain'' in equations, in the interest of clarity.}
\begin{eqnarray}
\text{information gain }(x_i) & = & \sum_j{p(y_j|x_i) \, \text{evidence}(y_j|x_i)} \nonumber \\
                      & = & \sum_j{p(y_j|x_i) \, \log_{10}{\frac{p(y_j|x_i)}{p(y_j)}}}.
\label{eqn:informationGain}
\end{eqnarray}
The scientific merit of the experimental result $x_i$, quantified by Eq.~\ref{eqn:informationGain} and motivated in terms of a weighted average of expected evidence, is the appropriate recasting of Eq.~\ref{eqn:eqn1} in terms of the information gained on the state of knowledge $Y$ from the experimental result $x_i$.  The information gain defined in Eq.~\ref{eqn:informationGain} is also sometimes referred to as information divergence, relative entropy, or the Kullback--Leibler divergence~\cite{KullbackLeibler} in the fields of probability theory and information theory, and the maximization of this utility function is referred to as Bayes d-optimal experimental design~\cite{BayesianExperimentalDesign}.   Intuitively, the information gain is a measure of how much our state of knowledge has changed as a result of obtaining the outcome $x_i$.  Eq.~\ref{eqn:informationGain} reduces to the previous Eq.~\ref{eqn:eqn1} when the experimental result $x_i$ uniquely specifies a corresponding state of knowledge $y_j$.  In this case $p(y_j|x_i)=1$, and Eq.~\ref{eqn:eqn1} quantifies our surprise at finding $y_j$ to be correct.  

Since the information gain is the appropriate measure of scientific merit of an experimental result $x_i$, the scientific merit of a proposed experiment should be quantified by the information that would be gained from each possible experimental outcome, weighted by the expectation $p(x_i)$ that each experimental outcome $x_i$ is obtained.  The information content $\Delta H$ in an experiment, defined earlier in Eq.~\ref{eqn:deltaInformationEntropy}, can be shown with a few lines of algebra~\footnote{
Proof: \\
$
\sum_i{p(x_i) \, \, \text{information gain }(x_i)} = \\
\sum_i{p(x_i) \, \sum_j{p(y_j|x_i) \, \text{evidence}(y_j|x_i) }} = \\
\sum_i{p(x_i) \, \sum_j{p(y_j|x_i) \, \log_{10}{\frac{p(y_j|x_i)}{p(y_j)}}}} = \\
\sum_{i,j}{p(x_i) \, p(y_j|x_i) \, \log_{10}{\frac{p(y_j|x_i)}{p(y_j)}}} = \\
\sum_{i,j}{p(x_i,y_j) \, \log_{10}{\frac{p(y_j|x_i)}{p(y_j)}}} = \\
\sum_{i,j}{p(x_i,y_j) \, [ \log_{10}{p(y_j|x_i)} - \log_{10}{p(y_j)} ] } = \\
- \sum_{i,j}{p(x_i,y_j) \log_{10}{p(y_j)}} + \sum_{i,j}{p(x_i,y_j) \log_{10}{p(y_j|x_i)}}=
- \sum_{j}{p(y_j) \log_{10}{p(y_j)}} - H(Y|X) = \\
H(Y) - H(Y|X) = \Delta H.
$
} to be equal to its expected information gain,
\begin{equation}
\Delta H=\sum_i{p(x_i) \, \text{information gain }(x_i)},
\end{equation}
as required for logical consistency of the concepts that have been introduced.

This section has introduced an additional level of sophistication beyond Sec.~\ref{sec:InformationEntropy:Basic}.  The elementary notion of surprisal in Eq.~\ref{eqn:eqn1}, quantifying the ({\em{a posteriori}}) scientific merit of an experimental result, is replaced by the information gain of Eq.~\ref{eqn:informationGain}.  The elementary notion of information content in Eq.~\ref{eqn:eqn2}, quantifying the ({\em{a priori}}) scientific merit of a proposed experiment, is replaced by Eq.~\ref{eqn:deltaInformationEntropy}.

\subsection{Synopsis}
\label{sec:Synopsis}

The essential thesis of this article is summarized in two sentences.
\begin{itemize}
\item The appropriate quantification of scientific merit of a proposed experiment or analysis (before it is performed and its outcome is known) is the reduction in information entropy the experiment or analysis is expected to provide, defined by Eq.~\ref{eqn:deltaInformationEntropy}.
\item The appropriate quantification of scientific merit of an experiment or analysis after the result is known is the information gained from the result, defined by Eq.~\ref{eqn:informationGain}.
\end{itemize}

\section{Examples}
\label{sec:Examples}

The specific figure of merit defined above can be used to quantify the scientific merit of any experiment or analysis.  This section illustrates the application of this figure of merit to particular experiments and analyses in the field of particle physics.

In these examples it will be clear that {\em{a priori}} estimates of the probability of possible outcomes must be made.  No prescription exists for choosing these {\em{a priori}} probabilities of possible outcomes.  The examples below will nonetheless demonstrate that (1) the same scientific conclusion is often reached for the full range of justifiable {\em{a priori}} probabilities, and (2) the formulation of the discussion in terms of the assignment of these {\em{a priori}} probabilities is frequently productive and enlightening.  The fact that there is not a well-developed literature to point to for the justification of these {\em{a priori}} probabilities emphasizes the fact that up to this point the importance of these probabilities has not been properly recognized in assessing the scientific merit of proposed and completed experiments.

\subsection{Proposed experiments}

For the purposes of the calculations below, the possibility of new physics being found at Tevatron Run II is taken to be 20\%, with 10\% in each of Run IIa and Run IIb, and with equal probability to be found in any of roughly $10^4$ potentially interesting kinematic distributions.  A more detailed justification of these numbers is provided in Appendix~\ref{sec:DiscoveryExpectationAppendix:Colliders}.  An observed discrepancy is assumed to contain sufficiently detailed information to reduce the space of qualitatively different possible explanations for the discrepancy by a factor of $10^2$.  These numbers will naturally be the subject of significant and productive debate in any real program review.  One of the purposes of the calculations below is to provide intuition for the way in which conclusions change with different probability assignments, and the extent to which they do not.

\begin{itemize}
\item Flipping a coin has two possible outcomes.  Let $x_1$ denote the experimental outcome in which the coin comes up heads, and let $x_2$ denote the experimental outcome in which the coin comes up tails.  Let $y_1$ denote the state of knowledge in which the Standard Model Higgs boson is known to exist, and let $y_2$ denote the state of knowledge in which the Standard Model Higgs boson is known not to exist.  Whether the coin comes up heads or tails tells us absolutely nothing about whether the Standard Model Higgs boson exists or not; hence $p(y_1|x_1)=p(y_1|x_2)=p(y_1)$, and $p(y_2|x_1)=p(y_2|x_2)=p(y_2)$.  In words, our degree of belief $p(y_1|x_1)$ that the Higgs boson exists given that the coin comes up heads is equal to our degree of belief $p(y_1|x_2)$ that the Higgs boson exists given that the coin comes up tails, which in turn is equal to our degree of belief $p(y_1)$ that the Higgs boson exists before performing the experiment.  In this case from Eq.~\ref{eqn:informationEntropy_posterior} the expected information entropy after performing the experiment $H(Y|X)= - \sum_{i,j}{p(x_i,y_j) \log_{10}{p(y_j|x_i)}}= - \sum_{j}{p(y_j) \log_{10}{p(y_j)}}=H(Y)$, the information entropy before performing the experiment, and the scientific merit of this experiment is $\Delta H=H(Y)-H(Y|X)=$~{\tt 0}.

This example illustrates that $\Delta H$ quantifies {\em{scientific}} merit (and not merely the number of possible experimental outcomes) through its treatment of qualitatively different states of scientific knowledge $Y$.  Experiments with varied possible outcomes $X$ that do not inform the probabilities we assign to the qualitative states of knowledge $Y$ have zero scientific merit.

\item The $\mu2e$ experiment~\footnote{This experiment is proposed to measure (or bound) the transition $\mu\rightarrow e$ in the electric field of an atomic nucleus.  The $\mu2e$ collaboration URL is currently maintained at \href{http://mu2e.fnal.gov}{http://mu2e.fnal.gov}.} has two possible outcomes.  Appendix~\ref{sec:DiscoveryExpectationAppendix:mu2e} argues that an excess of muon to electron conversions will be observed with an expectation of $\approx 18\%$.  With probability of $1-0.18=0.82$ results consistent with the Standard Model will be obtained.  In either case the experiment results in a state of knowledge with zero information entropy, so $H(Y|X)=0$.  The information content in this analysis is thus $\Delta H = H(Y)-H(Y|X) = H(Y)-0 = -(0.18 \times \log_{10}(0.18) + 0.82 \times \log_{10}(0.82)) \approx$~{\tt 2e-01}~\footnote{The style of scientific notation here and below (writing {\tt 2e-01} rather than $2\times10^{-1}$) is used to emphasize the exponent.}.

The most important factor for determining the scientific merit of this analysis is the estimated probability of $18\%$ of observing an excess of muon to electron conversions, which sets the scale for the computed scientific merit.  Note that for an experiment such as $\mu2e$ with binary outcome the scientific merit is maximal when the estimated probability for a signal is $50\%$.  

\item A search for single top has two possible outcomes.  As argued in Appendix~\ref{sec:DiscoveryExpectationAppendix:Colliders}, with probability $1-10^{-5}=0.99999$ single top will be found as predicted by the Standard Model, treating this analysis as just one of $10^4$ distributions that can be considered in Tevatron Run II.  With probability $10^{-5}$ single top will be found to not be consistent with the Standard Model prediction.   The expected reduction in information entropy from this analysis is thus $\Delta H = H(Y)-H(Y|X) = H(Y)-0 = -(10^{-5} \times \log_{10}(10^{-5}) + (1-10^{-5}) \log_{10}(1-10^{-5})) \approx$~{\tt 5e-05}.  

The dominant factor in determining the scientific merit of this analysis is seen to be the estimated probability of not observing single top as predicted by the Standard Model.  In this example, as in all others, the possibility of an outright experimental mistake is ignored.

\item A search for a gluino ($\tilde{g}$) has $\sim 101$ possible outcomes.  The probability for new physics to be found in Tevatron Run IIa is 10\%, with this probability divided equally among $10^4$ distributions.  The probability for new physics to be found in the particular distribution used in the gluino analysis is thus $10\%/10^4 = 10^{-5}$.  If new physics is found, the signal events will contain information that should narrow down the qualitatively different possible explanations for a discrepancy in this distribution by a factor of $10^2$.  That is, for every 100 qualitatively different plausible interpretations (\Bard\ stories~\cite{BardPRL:Knuteson:2006ha}) that could be proposed for seeing a discrepancy in the particular kinematic distribution considered, the characteristics of the observed events are expected to allow the elimination of 99 of them.  The probability that no new physics is found in this analysis is $1-10^{-5}$.  There are thus $10^2$ possible outcomes with probability $10^{-7}$ corresponding to $10^2$ possible different states of understanding in the event of a discovery, and one possible outcome with probability $1-10^{-5}=0.99999$ corresponding to no discovery.  The information content in this analysis is $\Delta H = H(Y)-H(Y|X) = H(Y)-0 = -( 10^2 \times 10^{-7} \times \log_{10}(10^{-7}) + (1-10^{-5}) \times \log_{10}(1-10^{-5})) \approx$~{\tt 7e-05}.  

The dominant factor in determining the scientific merit of this analysis is seen to be the estimated probability for discovery, which sets the scale of $10^{-5}$, adjusted slightly by the decimal logarithm of the estimated power of the analysis for discriminating among competing interpretations in the event of a discovery.  This analysis can be taken as a proxy for any similar dedicated Tevatron search.  

\item A search for the Standard Model Higgs boson at the Tevatron has (qualitatively) three possibilities.  The Standard Model Higgs boson exists with probability 95\%, and does not exist with probability 5\%.  If the Standard Model Higgs boson exists, it will be seen at the Tevatron with probability 10\%.  Possibilities are that the Standard Model Higgs boson exists and is seen at the Tevatron, with probability $p(\text{exists},\text{observed})=95\% \times 10\%=0.095$; that the Higgs boson exists but is not seen at the Tevatron, with probability $p(\text{exists},\text{!observed})=95\% \times 90\% = 0.855$; and that the Higgs boson does not exist and is not seen at the Tevatron, with probability $p(\text{!exists},\text{!observed})=5\%$.  
The entropy of the state of knowledge before the experiment is performed is $H(Y)=H(\text{exists?})=-(0.95 \times \log_{10}{0.95} + 0.05 \times \log_{10}{0.05}) \approx 0.086$~\footnote{To clarify notation:  $y_1=$~``\text{exists}'' refers to the state of knowledge in which the Higgs boson is known to exist; $y_2=$~``\text{!exists}'' refers to the state of knowledge in which the Higgs boson is known not to exist; and $Y=$~``\text{exists?}'' refers to the two possible states of knowledge $y_1$ and $y_2$.  Similarly, $x_1=$~``\text{observed}'' refers to the experimental outcome in which the Higgs boson is observed; $x_2=$~``\text{!observed}'' refers to the experimental outcome in which the Higgs boson is not observed; and $X=$~``\text{observed?}'' refers to the two possible experimental outcomes $x_1$ and $x_2$.}.
Unlike the previous examples, in which the experiment results in an information state with zero entropy, in the search for the Higgs boson at the Tevatron we may be left in an information state with non-zero entropy.  The scientific merit is the expected reduction in entropy $\Delta H = H(Y) - H(Y|X) = H(\text{exists?})-H(\text{exists?}|\text{observed?})$ that will be provided by the experiment, where
$H(\text{exists?}|\text{observed?})= -(
p(\text{exists},\text{observed})\times\log_{10}{p(\text{exists}|\text{observed})}+
p(\text{exists},\text{!observed})\times\log_{10}{p(\text{exists}|\text{!observed})}+
p(\text{!exists},\text{!observed})\times\log_{10}{p(\text{!exists}|\text{!observed})}
) \approx -(
0.095\times\log_{10}(1)+
0.855\times\log_{10}(0.945)+
0.05 \times\log_{10}(0.055)
) \approx 0.084$.  
The scientific merit is thus calculated to be $\Delta H = H(Y) - H(Y|X) \approx 0.086-0.084=$~{\tt 2e-03}.

The important factors for determining the scientific merit of this analysis are the estimated probability of 95\% of observing the Higgs, which sets the scale of a few $\times 10^{-2}$, and the estimated probability of 10\% that the experiment will provide an unambiguous answer, which reduces the scientific merit to a few $\times 10^{-3}$.

\item Rewinding back to the year 2003, the Tevatron $B_s$ mixing measurement with 1~fb$^{-1}$ of integrated luminosity similarly has (qualitatively) three possibilities.  With initial results from the $B$ factories already in hand, the Standard Model CKM picture will be insufficient to accommodate the Tevatron measurement of $\Delta m_s$ with probability $10^{-5}$, as argued in Appendix~\ref{sec:DiscoveryExpectationAppendix:Colliders}.  The projected sensitivity of the CDF $B_s$ mixing analysis with 1~fb$^{-1}$ is $\Delta m_s = 30$~ps$^{-1}$.  The Standard Model can accommodate a value of $\Delta m_s$ up to and including 30~ps$^{-1}$ at 95\% confidence level.  If the CKM picture is sufficient, with probability 95\% a measurement of $\Delta m_s$ will made with $\Delta m_s <30$~ps$^{-1}$, confirming that the Standard Model CKM picture is indeed sufficient.  Possibilities are that the CKM picture is sufficient and $B_s$ mixing will be observed, with probability $p(\text{ckm},\text{observed})=0.99999 \times 95\% \approx 0.94999$; that the CKM picture is sufficient but $\Delta m_s$ is sufficiently large that $B_s$ mixing is not seen at the Tevatron, with probability $p(\text{ckm},\text{!observed})=0.99999 \times 5\% \approx 0.0499995$; and that the CKM picture is insufficient and $B_s$ mixing is not seen at the Tevatron, with probability $p(\text{!ckm},\text{!observed})=10^{-5}$.  The entropy of the state of knowledge before the experiment is performed is $H(Y)=H(\text{ckm?})=-(0.99999 \times \log_{10}{0.99999} + 0.00001 \times \log_{10}{0.00001}) \approx 5\times10^{-5}$~\footnote{To again clarify notation:  $y_1=$~``\text{ckm}'' refers to the state of knowledge in which the CKM picture is sufficient; $y_2=$~``\text{!ckm}'' refers to the state of knowledge in which the CKM picture is insufficient; and $Y=$~``\text{ckm?}'' refers to the two possible states of knowledge $y_1$ and $y_2$.  Similarly, $x_1=$~``\text{observed}'' refers to the experimental outcome in which $B_s$ mixing is observed, and $\Delta m_s$ is found to be $<30$~ps$^{-1}$; $x_2=$~``\text{!observed}'' refers to the experimental outcome in which $B_s$ mixing is not observed, and $\Delta m_s$ is found to be $>30$~ps$^{-1}$; and $X=$~``\text{observed?}'' refers to the two possible experimental outcomes $x_1$ and $x_2$.}.
As in the previous Higgs example, this experiment may result in an information state with non-zero entropy.  The appropriate figure of merit is the entropy reduction $\Delta H = H(Y)-H(Y|X) = H(\text{ckm?})-H(\text{ckm?}|\text{observed?})$, where
$H(\text{ckm?}|\text{observed?})= -(
p(\text{ckm},\text{observed})\times\log_{10}{p(\text{ckm}|\text{observed})}+
p(\text{ckm},\text{!observed})\times\log_{10}{p(\text{ckm}|\text{!observed})}+
p(\text{!ckm},\text{!observed})\times\log_{10}{p(\text{!ckm}|\text{!observed})}
) \approx -(
0.94999\times\log_{10}(1)+
0.0499995\times\log_{10}(0.9998)+
10^{-5} \times\log_{10}(0.0002)
) \approx 4\times10^{-5}$. 
The scientific merit is the reduction in information entropy expected to be provided by the experiment, calculated to be $\Delta H =H(Y)-H(Y|X) \approx 5\times10^{-5}-4\times10^{-5}=$~{\tt 1e-05}.

The most important factor for determining the scientific merit of this analysis is the estimated probability of $10^{-5}$ of not observing $B_s$ mixing, which sets the scale of a few $\times 10^{-5}$.  Unfortunately the sensitivity of this analysis is such that it is unable to cash in on the (sur)prize of providing conclusive evidence that the CKM picture is insufficient.  It is instructive to see how this limited sensitivity (resulting in the number of 5\% in the calculation above) propagates through the calculation into a relatively large value for $H(Y|X)$, resulting in an expected information entropy $H(Y|X)=4\times10^{-5}$ after the measurement that is comparable to the information entropy $H(Y)=5\times10^{-5}$ before the measurement, leaving an expected information gain of only $\Delta H = 1\times10^{-5}$.

\item A global analysis of all Tevatron Run II data will discover nothing with a probability of 80\%.  With probability 20\% a discovery will be made in one of $10^4$ distributions considered, with $10^2$ qualitatively different possible states of understanding resulting from a discrepancy observed in any of these distributions.  In the event of a discovery, there are thus $10^6$ possible states of understanding, each with probability $2\times 10^{-7}$.  The information content in this analysis is thus $\Delta H = H(Y)-0 = -(10^6 \times (2\times 10^{-7}) \times \log_{10}(2\times 10^{-7}) + 80\% \times \log_{10}(80\%)) =$~{\tt 1.4}.

The dominant factors influencing the scientific merit of this analysis are the estimated probability of $20\%$ of discovering new physics at Tevatron Run II, which sets the scale of $0.2$, and the decimal logarithm of the power of the analysis to discriminate among possible interpretations, which provides an additional factor of seven.

\item Integrating over all analyses that will be performed at the CERN Large Hadron Collider in the next five years, a significant discovery will be made with a probability of 90\%.  The space of new physics models is so large that there are $\sim 10^8$ qualitatively distinct states of possible understanding accessible in the event of a discovery.  With probability of 10\% a Standard Model Higgs boson will be seen and nothing else.  The information content in the entire LHC is thus $\Delta H = H(Y)-0 = -(10^8 \times (90\% \times 10^{-8}) \times \log_{10}(90\% \times 10^{-8}) + 10\% \times \log_{10}(10\%)) =$~{\tt 7.4}.

In this case the $O(1)$ discovery potential sets the scale for LHC's scientific merit, and the power of the direct production of new states to distinguish among competing interpretations provides an additional factor of seven.

\end{itemize}

The assignment of {\em{a priori}} probabilities above is certainly open for debate, and the reader may object that the problem of quantifying an experiment's scientific merit has simply been reformulated in terms of the estimation of the probabilities of possible experimental outcomes.  At worst, this reformulation significantly changes and focuses the discussion.  In most cases, the scientific conclusion drawn from this reformulation is robust against the variation of these estimated probabilities within their justifiable range.  In cases where the scientific conclusion is not robust against such variation, this framework points out precisely what factors are important.  The fact that there is not a well-developed literature to point to for the justification of the above estimated probabilities emphasizes the fact that until now these estimated probabilities have not been regarded as {\em{the}} important ingredients in assessing the scientific merit of proposed and completed experiments.

\begin{table}
\begin{tabular}{llll}
\multicolumn{1}{c}{Analysis} & 
\multicolumn{1}{c}{Merit\hspace{0.3cm}} & 
\multicolumn{1}{c}{Cost\hspace{0.3cm}} & 
\multicolumn{1}{c}{Bang per buck} \\
\multicolumn{1}{c}{} & 
\multicolumn{1}{c}{} &
\multicolumn{1}{c}{(M\$)} & 
\multicolumn{1}{c}{(Merit per M\$)} \\ \hline
LHC, all searches & {\tt 7} & \hspace{0.1cm}{\tt 5e+03}\hspace{0.2cm} & \hspace{0.6cm}{\tt 1e-03} \\
Tevatron II global search & {\tt 1.4} & \hspace{0.1cm}{\tt 3e-01} & \hspace{0.6cm}{\tt 5} \\
$\mu2e$ & {\tt 2e-01} & \hspace{0.1cm}{\tt 1e+02} & \hspace{0.6cm}{\tt 2e-03} \\
Tevatron Higgs search & {\tt 2e-03} & \hspace{0.1cm}{\tt 1e+01} & \hspace{0.6cm}{\tt 2e-04} \\
$B_s$ mixing & {\tt 1e-05} & \hspace{0.1cm}{\tt 1e+01} & \hspace{0.6cm}{\tt 1e-06} \\
$\tilde{g}$ search & {\tt 7e-05} & \hspace{0.1cm}{\tt 1e-01} & \hspace{0.6cm}{\tt 7e-04} \\
single top search & {\tt 5e-05} & \hspace{0.1cm}{\tt 5} & \hspace{0.6cm}{\tt 1e-05} \\
flipping a coin & {\tt 0} & \hspace{0.1cm}{\tt 1e-07} & \hspace{0.6cm}{\tt 0} \\
\end{tabular}
\caption{The scientific merit of selected experiments and analyses, calculated in terms of expected information content $\Delta H$ with decimal logarithm as described in the text, and ordered according to decreasing merit.  Estimated incremental cost for each experiment or analysis is provided in units of millions of U.S.~dollars.  Scientific ``bang per buck'' is obtained by dividing the calculated merit by the estimated incremental cost.\label{tbl:scientificMeritOfProposedExperiments}}
\end{table}

The scientific merit of the proposed experiments and analyses considered above can in this way be quantified and ordered, as in Table~\ref{tbl:scientificMeritOfProposedExperiments}.  Also included in this table is an estimated incremental cost for performing the experiment or analysis, which is subject as always to some ambiguity in the choice of accounting~\footnote{Such an estimation of incremental cost is inherently problematic due to ambiguities in accounting.  Incremental cost for analyses within ``existing'' experiments is taken to be given by the total number of person-years required for the analysis, at an estimated cost of $10^2$~k\$ per person-year.  Incremental costs for analyses representing the primary justification for an experiment's construction are taken to be the total cost of the experimental apparatus.}.   Dividing the scientific merit by the estimated cost provides a ``scientific bang per buck,'' in units of information content per million dollars.  Since information entropy has the desired behavior under addition (in the sense that if two experiments together provide the same information as a third experiment, the sum of the merits of the first two experiments equals the merit of the third), the scientific merit per unit cost is appropriate for judging the most cost effective allocation of limited resources, and for comparing experiments of significantly different scales.  This ``scientific bang per buck'' provides a quantitative framework for advocates of experiments of all sizes to argue their cases on even footing, and for review panels to assess such cases on even footing.

In this context, the value of making data publicly available (such as through \Quaero~\cite{QuaeroPRL:Abazov:2001ny2,QuaeroH1:Caron:2006fg}) is to reduce the cost of dedicated analysis (such as the $\tilde{g}$ search in Table~\ref{tbl:scientificMeritOfProposedExperiments}) from 100~k\$ (corresponding to two years of a graduate student's time) to 100~\$ (corresponding to two hours of a graduate student's time), increasing the scientific bang per buck for such a search by three orders of magnitude.

Table~\ref{tbl:scientificMeritOfProposedExperiments} is particularly interesting and potentially valuable to the extent that its conclusions are at odds with accepted wisdom of the relative scientific merit of the experiments and analyses listed.  If indeed the figure of merit proposed here is a suitable measure of the scientific merit of an experiment or analysis, then accepted wisdom can be dangerously misleading.  If accepted wisdom is correct, then there should exist an alternative and justifiably more appropriate quantifiable figure of merit than the one proposed in this article.  If the latter, this article should help motivate its development.

\subsection{Completed experiments}

The figure of merit suggested for assessing the scientific merit of a completed experiment is the information gained from the result obtained, quantified by Eq.~\ref{eqn:informationGain}.  This section illustrates the application of this figure of merit to historical, recent, and hypothetical future experimental results in our field.

In these examples, as in those in the previous section, it will be clear that estimates of the probability of possible outcomes must be made.  For completed experiments, like for future experiments, the relevant probabilities are those that quantify our expectation of probable outcomes before the execution of the experiment.  Here as before, no prescription exists for choosing these {\em{a priori}} probabilities of possible outcomes.  The examples below will nonetheless demonstrate again that (1) the same scientific conclusion is often reached for the full range of justifiable {\em{a priori}} probabilities, and (2) the formulation of the discussion in terms of the assignment of these {\em{a priori}} probabilities is frequently productive and enlightening.  Provided historical references, although not explicitly quantifying expectation, nonetheless give a flavor of the rhetoric of the time that can be used to justify reasonable {\em{a priori}} probabilities.

\begin{itemize}
\item Returning to the example of the flipped coin, let $x_1$ denote the experimental outcome in which the coin comes up heads, $x_2$ the experimental outcome in which the coin comes up tails, $y_1$ the state of knowledge in which the Standard Model Higgs boson is known to exist, and $y_2$ the state of knowledge in which the Standard Model Higgs boson is known not to exist.  As before, $p(y_1|x_1)=p(y_1|x_2)=p(y_1)$, and $p(y_2|x_1)=p(y_2|x_2)=p(y_2)$.  Assume the coin comes up heads:  the experimental outcome $x_1$ is realized.  

In this case from Eq.~\ref{eqn:informationGain} the information gain of this outcome is $-\sum_j{p(y_j|x_1) \, \log_{10}{\frac{p(y_j)}{p(y_j|x_1)}}} = -\sum_j{p(y_j) \, \log_{10}{\frac{p(y_j)}{p(y_j)}}} = -\sum_j{p(y_j) \, \log_{10}{1}} = 0$.  The scientific merit of this experimental outcome, quantified by its information gain, is~{\tt 0}.

\item Before the $W$ and $Z$ bosons were discovered~\cite{UA1Wdiscovery:Arnison:1983rp,UA2Wdiscovery:Banner:1983jy,UA1Zdiscovery:Arnison:1983mk,UA2Zdiscovery:Bagnaia:1983zx}, these electroweak gauge bosons were expected to exist with a degree of confidence sufficiently high that the S$p\bar{p}$S was specifically designed and built for their discovery~\cite{SppbarSjustification:Rubbia:1979wm}.  In the late 1970's, the $W$ and $Z$ bosons were assumed to exist with a probability of 95\%~\cite{Taubes}.  The information gain of the discovery of these gauge bosons in 1983 was thus $-\log_{10}(95\%)\approx$~{\tt 0.02}.
\item Before the top quark was discovered at Fermilab in 1995~\cite{CDFTopDiscovery:Abe:1995hr1,D0TopDiscovery:Abachi:1995iq1}, it was assumed to exist with a degree of confidence sufficiently high that the Tevatron to a great extent was specifically designed and built for this discovery~\cite{TevatronTopQuark:Lederman:1988hy}.  In the late 1980's, the top quark was assumed to exist with a probability of 95\%~\cite{TevatronTopQuarkExpectation:Shochet:1989dq}.  The information gain of the discovery of the top quark in 1995 was thus $-\log_{10}(95\%)\approx$~{\tt 0.02}.
\item In 1990, it was believed with a probability of $30\%$ that new electroweak scale physics (such as supersymmetry) would appear in the running of Tevatron Run I or LEP\,2.  The suprisal that no new physics was discovered at Tevatron Run I or LEP\,2 was thus $-\log_{10}(1-30\%)\approx$~{\tt 0.15}~\footnote{Although nothing beyond the Standard Model was discovered at Tevatron Run I or LEP\,2, it should be noted that the number of published distributions comparing data to Standard Model prediction in both of these collider runs represents only roughly 1\% of the distributions that might plausibly be expected to contain the first hint of new physics.  For the purpose of this discussion (but contrary to fact), we assume the remaining distributions were all carefully considered and no anomalous effect was observed.}.
\item The discovery of the $J/\Psi$~\cite{TingCharmDiscovery:Aubert:1974js,RichterCharmDiscovery:Augustin:1974xw} was largely unanticipated; in 1973 few professional physicists took the notion of physical quarks seriously~\cite{Pickering}.  The expectation that experiments at Brookhaven and SLAC would produce evidence of a narrow resonance confirming the physicality of the quark picture was small, corresponding to a probability of $\sim 10^{-2}$.  The information gain of the discovery of the $J/\Psi$ in November of 1974 was thus $-\log_{10}(10^{-2})=$~{\tt 2}.
\item The discovery of the $\tau$ lepton~\cite{TauDiscovery1:Perl:1975bf,TauDiscovery2:Perl:1976rz,TauDiscovery3:Perl:1977se} was also quite unanticipated; in the early 1970's experimental results fit nicely into a framework with only two generations of fermions.  Indeed, there was less theoretical motivation for expecting the $\tau$ lepton than expecting to see the $J/\Psi$~\cite{Feldman:1992vk}.  The expectation that a third generation would be discovered thus corresponded to a probability even less than the number used above for the $J/\Psi$ discovery.  If the {\em{a priori}} probability of a SLAC experiment producing evidence for a $\tau$ lepton in 1977 is taken to be $10^{-3}$, then the information gain of this experimental result is $-\log_{10}(10^{-3})=$~{\tt 3}.
\item By 1977, the existence of a third generation of leptons was established~\cite{TauDiscovery3:Perl:1977se}.  The expectation within the community at this time that a $b$ quark would be observed can be estimated to correspond to a probability of $\sim 30\%$~\cite{UpsilonDiscoveryExpectation:Cahn:1977dv}.  The discovery of the $\Upsilon$~\cite{LedermanUpsilonDiscovery:Herb:1977ek} thus represents an experimental result with scientific merit equal to $-\log_{10}(30\%)\approx$~{\tt 0.5}.
\item The determination of $\Delta m_s$ has recently been improved at the Tevatron with 1~fb$^{-1}$ of integrated luminosity through an analysis of $B_s$ mixing~\cite{CDFBsMixing:Abulencia:2006ze}.  Before the observation of $B_s$ mixing, the CKM picture was expected to provide an inadequate description of this phenomenon with a probability of $10^{-5}$.  Quantified in terms of information gain, the scientific merit of this experimental result is $-\log_{10}(0.99999)=$~{\tt 4e-06}.
\item A search for a gluino ($\tilde{g}$) has been performed at the Tevatron, with no evidence of new physics found.  From the previous section, the probability of finding a gluino in this search was $10^{-5}$.  The scientific merit of the result of this search, in which no evidence of physics beyond the Standard Model is found, is quantified by a information gain of $-\log_{10}(0.99999)=$~{\tt 4e-06}.
\item A global analysis of CDF Run IIa high-$p_T$ data has recently been performed~\cite{VistaCDFPRL:Collaboration:2007kp,VistaCDFPRD:Collaboration:2007dg}, and no evidence of new physics has been found.  Using numbers from the previous section, the probability that no evidence of new physics would be found in this analysis was $90\%$.  The scientific merit of this result is thus equal to $-\log_{10}(90\%)=$~{\tt 0.05}.
\item Fast-forwarding in time, consider the scientific merit of a null result obtained in a global search of Tevatron Run IIb high-$p_T$ data.  Using numbers from the previous section, the probability that no evidence of new physics would be found in these data is $90\%$.  The scientific merit of such a result would thus be equal to $-\log_{10}(90\%)=$~{\tt 0.05}.

The total scientific merit of a null result being obtained in global searches at Tevatron Run IIa and Tevatron Run IIb would be $0.05+0.05=$ {\tt 0.10}.

\item Fast-forwarding in time, consider the scientific merit of the discovery of a Standard Model Higgs boson.  The overwhelming expectation of the field is that a Standard Model Higgs boson exists, with a probability taken to be $\sim 95\%$.  The information gain of obtaining this experimental result is $-\log_{10}(95\%)=$~{\tt 0.02}.  If the LHC instead rules out the possibility of the existence of a Standard Model like Higgs boson, the scientific merit of such a result would be significantly greater, with information gain equal to $-\log_{10}(5\%)=$~{\tt 1.3}.
\item Fast-forwarding in time, consider the scientific merit of the discovery of Standard Model single top production.  The overwhelming expectation is that the top quark is produced singly in $p\bar{p}$ collisions as predicted by the Standard Model.  Taking the probability that Nature does not singly produce top quarks from the previous section to be $10^{-5}$, as argued in Appendix~\ref{sec:DiscoveryExpectationAppendix:Colliders}, the scientific merit of the discovery of single top production is quantified by a information gain of $-\log_{10}(0.99999)=$~{\tt 4e-06}.
\end{itemize}

The assignment of probabilities of possible outcomes above is certainly open for debate.  The claim is made that the physics discussion of these probabilities forms a particularly focused, fruitful, and constructive framework for assessing the scientific merit of experimental results.  

\begin{table}
\begin{tabular}{llll}
\multicolumn{1}{c}{Result} & 
\multicolumn{1}{c}{\hspace{0.15cm}Merit\hspace{0.15cm}} & 
\multicolumn{1}{c}{\hspace{0.15cm}Cost\hspace{0.15cm}} & 
\multicolumn{1}{c}{Bang per buck} \\
\multicolumn{1}{c}{} & 
\multicolumn{1}{c}{\hspace{0.1cm}} & 
\multicolumn{1}{c}{\hspace{0.1cm}(M\$)} & 
\multicolumn{1}{c}{(Merit per M\$)} \\ \hline
$\tau$ discovery & \hspace{0.2cm}{\tt 3} & \hspace{0.1cm}{\tt 6e-01} & \hspace{0.6cm}{\tt 5e+00} \\
$J/\Psi$ discovery & \hspace{0.2cm}{\tt 2} & \hspace{0.1cm}{\tt 1e+01} & \hspace{0.6cm}{\tt 2e-01} \\
there is no Higgs$\dagger$ & \hspace{0.2cm}{\tt 1.3} & \hspace{0.1cm}{\tt 5e+03} & \hspace{0.6cm}{\tt 3e-04} \\
$\Upsilon$ discovery & \hspace{0.2cm}{\tt 5e-01} & \hspace{0.1cm}{\tt 1} & \hspace{0.6cm}{\tt 5e-01} \\
null Tevatron I + LEP\,2 & \hspace{0.2cm}{\tt 2e-01} & \hspace{0.1cm}{\tt 3e+03} & \hspace{0.6cm}{\tt 6e-05} \\
global null Tevatron IIa  & \hspace{0.2cm}{\tt 5e-02} & \hspace{0.1cm}{\tt 3e-01} & \hspace{0.6cm}{\tt 2e-01} \\
global null Tevatron IIb$\dagger$  & \hspace{0.2cm}{\tt 5e-02} & \hspace{0.1cm}{\tt 3e-01} & \hspace{0.6cm}{\tt 2e-01} \\
$W$ and $Z$ discoveries & \hspace{0.2cm}{\tt 2e-02} & \hspace{0.1cm}{\tt 5e+02} & \hspace{0.6cm}{\tt 4e-05} \\
top quark discovery & \hspace{0.2cm}{\tt 2e-02} & \hspace{0.1cm}{\tt 5e+01} & \hspace{0.6cm}{\tt 4e-04} \\
Higgs discovery$\dagger$ & \hspace{0.2cm}{\tt 2e-02} & \hspace{0.1cm}{\tt 5e+03} & \hspace{0.6cm}{\tt 4e-06} \\
$B_s$ mixing observation & \hspace{0.2cm}{\tt 4e-06} & \hspace{0.1cm}{\tt 1e+01} & \hspace{0.6cm}{\tt 4e-07} \\
$\tilde{g}$ search & \hspace{0.2cm}{\tt 4e-06} & \hspace{0.1cm}{\tt 1e-01} & \hspace{0.6cm}{\tt 4e-05} \\
single top discovery$\dagger$ & \hspace{0.2cm}{\tt 4e-06} & \hspace{0.1cm}{\tt 5} & \hspace{0.6cm}{\tt 4e-06} \\
coin comes up heads & \hspace{0.2cm}{\tt 0} & \hspace{0.1cm}{\tt 1e-07} & \hspace{0.6cm}{\tt 0} 
\end{tabular}
\caption{Table of the scientific merit of selected historical, recent, and hypothetical future ($\dagger$) experimental results, calculated in terms of information gain with decimal logarithm as described in the text, and ordered according to decreasing merit.  Estimated incremental cost for each experiment or analysis is provided in units of millions of U.S.~dollars.  Scientific ``bang per buck'' is obtained by dividing the calculated merit by the estimated incremental cost.\label{tbl:scientificMeritOfExperimentalResults}}
\end{table}

The scientific merit of the experimental results considered above can thus be quantified and ordered, as in Table~\ref{tbl:scientificMeritOfExperimentalResults}.  Also included in this table is an estimated incremental cost for performing the experiment or analysis, which is subject as always to some ambiguity in the choice of accounting.  Dividing the scientific merit by the estimated cost provides a ``scientific bang per buck,'' in units of information content per million dollars.

Table~\ref{tbl:scientificMeritOfExperimentalResults} is particularly interesting and potentially valuable to the extent that its conclusions are at odds with accepted wisdom of the relative scientific merit of the experiments and analyses listed.  Note that the scientific merit, quantified in terms of the information gain, obeys the desired property of addition, such that if two experiments together come to the same conclusion as a third experiment, then the sum of the scientific merits of the first two experimental results is equal to the scientific merit of the third result.  The scientific merit of the discovery of the $\tau$ lepton is seen to equal the total scientific merit of all other present and past experimental results listed in Table~\ref{tbl:scientificMeritOfExperimentalResults}, a conclusion that is robust against reasonable variation in estimated prior probabilities.  The scientific merit of a Tevatron global search is seen to be $10^{4}$ times that of a dedicated, model-dependent Tevatron search, consistent with the relative scope of data space analyzed in each case.  The scientific merit of the combined null result of Tevatron I and LEP\,2 is measured to be twice the combined scientific merit of the discovery of the $W$ boson, the discovery of the $Z$ boson, the discovery of the top quark, and a future Higgs boson discovery.  The scientific merit of not finding a Standard Model Higgs boson at the LHC is nearly two orders of magnitude larger than the scientific merit of discovering the Standard Model Higgs boson.  The scientific merits of the ``discoveries'' of the $W$ boson, $Z$ boson, top quark, Standard Model Higgs, $B_s$ mixing, and single top production range from two to five orders magnitude less than the scientific merits of the ``null results'' that would have corresponded to a non-observation.   

\section{Summary}

The choice of a reasonable quantitative figure of merit for assessing the scientific merit of proposed experiments can inform and focus program review and accompanying decisions of resource allocation in many subfields of the physical sciences.  The related choice of a reasonable figure of merit for assessing the scientific merit of any particular experimental result would inform the evaluation of those organizations and individuals responsible for the production of the result.

This article advocates that the amount of information an experiment is expected to provide is the appropriate figure of merit for assessing the scientific merit of any experiment whose result is not yet known.  The amount of information an experiment is expected to provide is a concept with a well developed theory that is nearly sixty years old, quantified by a change in information entropy $\Delta H$ in the context of information theory.

This article advocates that the extent to which an experimental result surprises provides the appropriate figure of merit for assessing the scientific merit of any specific experimental result.  The extent to which an experimental result surprises is also an elementary notion in the context of information theory, quantified by the result's information gain.

Use of information content or information gain to evaluate the scientific merit of experiments requires the estimation of the probabilities of qualitatively different outcomes, and the reader may object that the problem of quantifying an experiment's scientific merit has simply been reformulated in terms of the estimation of the probabilities of possible experimental outcomes.  At worst, this reformulation significantly changes and focuses the discussion.   The fact that there is not a well-developed literature to point to for the justification of these {\em{a priori}} probabilities emphasizes the fact that until now the importance of these probabilities has not been properly recognized in assessing the scientific merit of proposed and completed experiments.  In most cases, the scientific conclusion drawn from this reformulation is robust against the variation of these estimated probabilities within their justifiable range.  From Tables~\ref{tbl:scientificMeritOfProposedExperiments} and~\ref{tbl:scientificMeritOfExperimentalResults}, it is seen that the scientific merit per incremental unit cost ranges over five orders of magnitude for the representative experiments and analyses considered.

The reader may object to the very idea of constructing an explicit figure of merit to quantify the scientific merit of experiments.  Such a reader misses the point that this is done (implicitly, if not explicitly) every time a decision of resource allocation is made.  It is surely in the field's best interest for such evaluations to be made in the sharpest, most open, most quantifiable, and scientifically best motivated framework possible.  

The reader may object to the specific figure of merit advocated in this article, particularly if his or her experiment scores poorly.  Such a reader is challenged to find a scientifically better motivated figure of merit.

Presumably nearly all readers will take issue with a few of the numbers in the examples.  Readers should use these to assess for themselves the claims made at the beginning of Sec.~\ref{sec:Examples}: (1) the same scientific conclusion is often reached for the full range of justifiable {\em{a priori}} probabilities, and (2) the formulation of the discussion in terms of the assignment of these {\em{a priori}} probabilities is frequently productive and enlightening.

In the context of the scientific merit advocated in this article, the point of making frontier energy collider data publicly available (such as with \Quaero~\cite{QuaeroPRL:Abazov:2001ny2,QuaeroH1:Caron:2006fg}) is to reduce the cost of an analysis like the $\tilde{g}$ search in Tables~\ref{tbl:scientificMeritOfProposedExperiments} and~\ref{tbl:scientificMeritOfExperimentalResults} from $\sim$\$100,000 to $\sim$\$100, increasing the scientific bang per buck of such an analysis by three orders of magnitude.

The figure of merit described here may provide a novel and useful quantitative framework within which (HEPAP, P5, NSF, and DoE) decisions about future resource allocation can be made.

\acknowledgments

The quantitative measure of experimental scientific merit proposed in this article grew through a thread of conversation held largely in the third week of September 2006.  Notable participants in this discussion include Steve Mrenna, Georgios Choudalakis, and Conor Henderson.  Bill Ashmanskas, Dan Marlow, Larry Rosenson, and Peter Fisher provided valuable feedback.  Discussions with Andre de Gouvea led to the assessment of the scientific merit of the $\mu2e$ experiment described in Appendix~\ref{sec:DiscoveryExpectationAppendix:mu2e}.

\appendix

\section{Derivation of expectation}
\label{sec:DiscoveryExpectationAppendix}

This appendix provides arguments for the numbers used to quantify expectation in the main text of this article.  The quantitative measure introduced in this article is sufficiently novel that the field so far has little practice with the type of argument required; reference to existing literature is in most cases insufficient.  The quantitative arguments constructed below are almost certainly also insufficient in some respects, but are provided as a starting point for further improvement and discussion.

\subsection{LEP, Tevatron, and LHC} 
\label{sec:DiscoveryExpectationAppendix:Colliders}

This section argues the discovery probabilities of LEP, the Tevatron, and the Large Hadron Collider.  A prior is constructed to represent our expectation in 1990 that the first sign of physics beyond the Standard Model will appear at or slightly above the electroweak scale.  This prior is then evolved through the experimental realities of Tevatron Run I and LEP\,2 to update expectation to the year 2000.  This expectation is then evolved through the experimental reality of Tevatron Run IIa and a hypothetical null result in Tevatron Run IIb to determine what fraction of expectation from 2000 is left at the end of Tevatron running in 2010.  The fraction of expectation from 1990 that has been eliminated by 2000, 2010, and 2020 (corresponding to the discovery probabilities of LEP, the Tevatron, and the LHC) is determined.

\begin{figure*}
\begin{tabular}{ccc}
\includegraphics[width=2.1in]{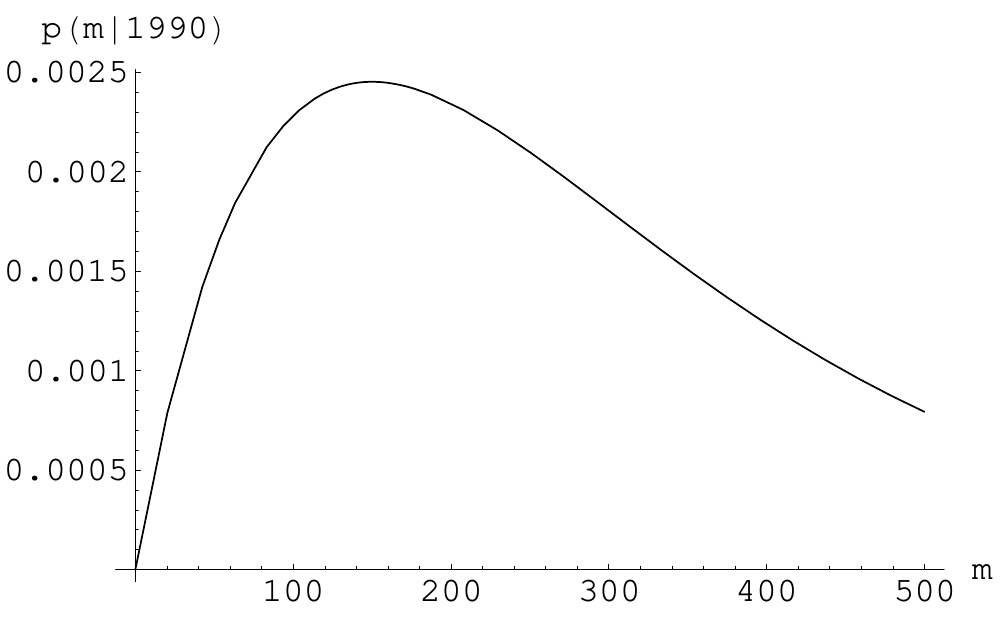} &
\includegraphics[width=2.1in]{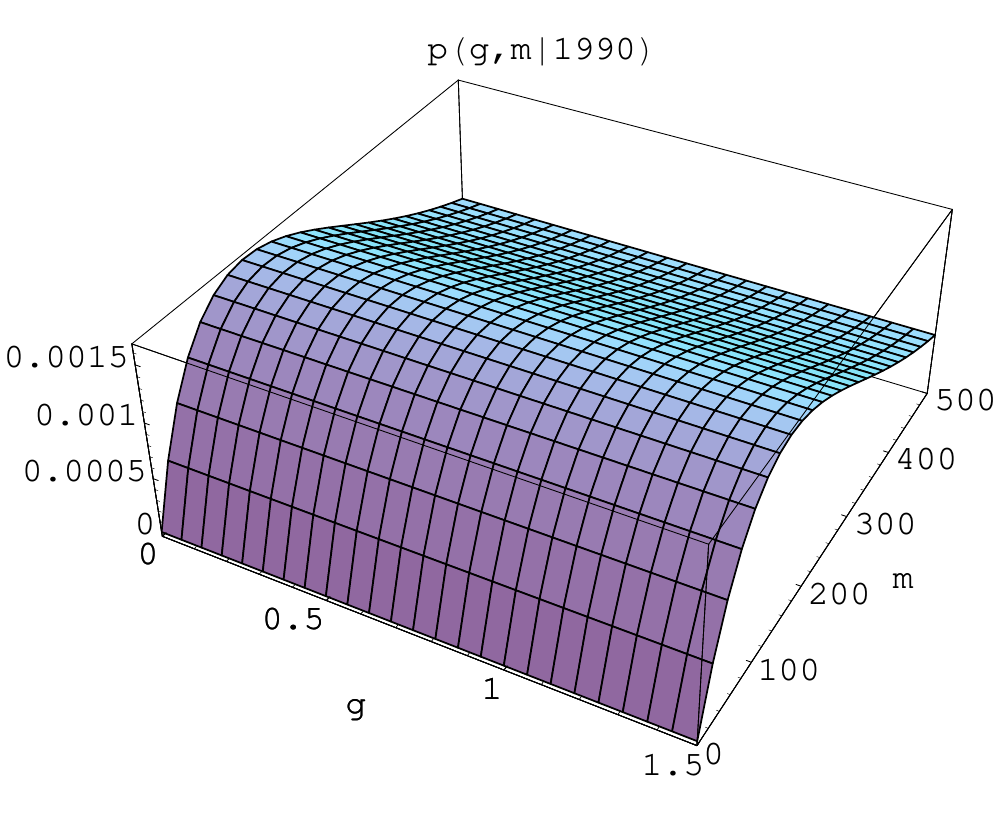} &
\includegraphics[width=2.1in]{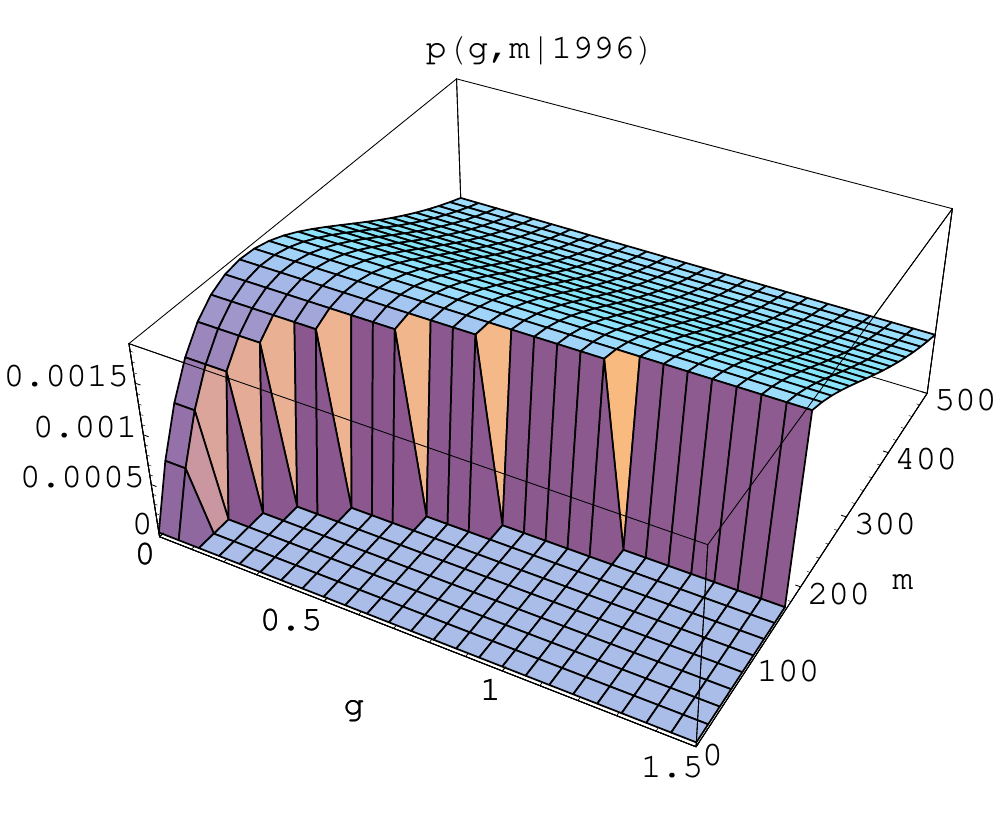} \\ 
\includegraphics[width=2.1in]{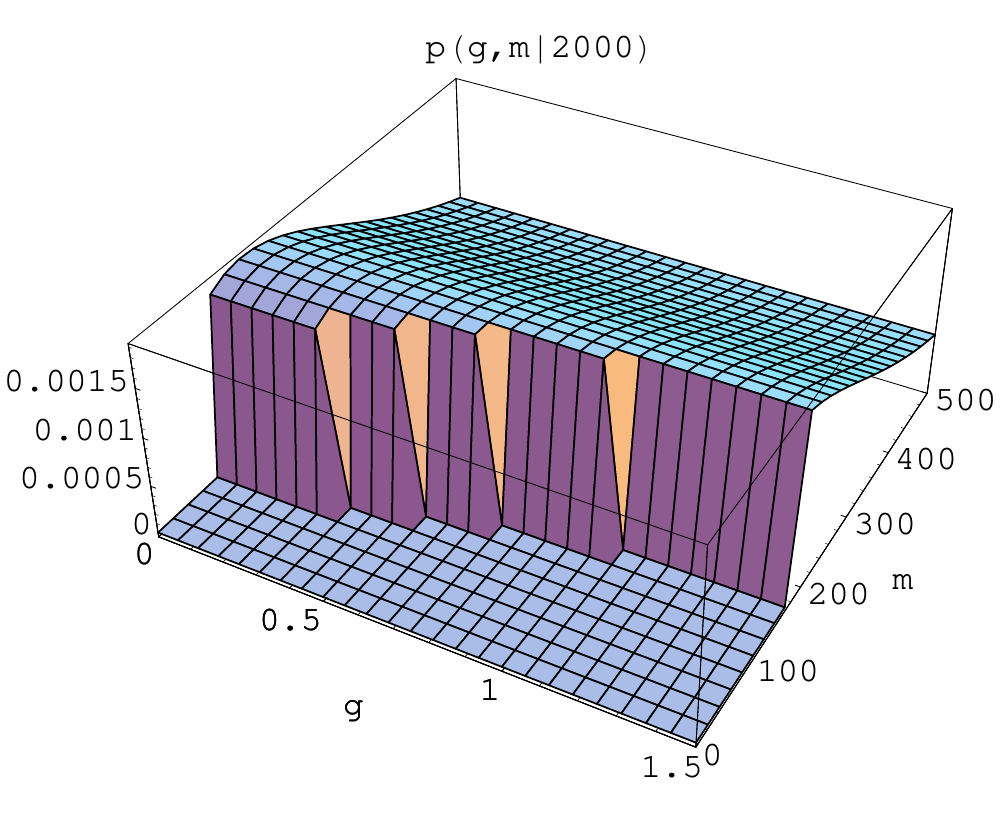} &
\includegraphics[width=2.1in]{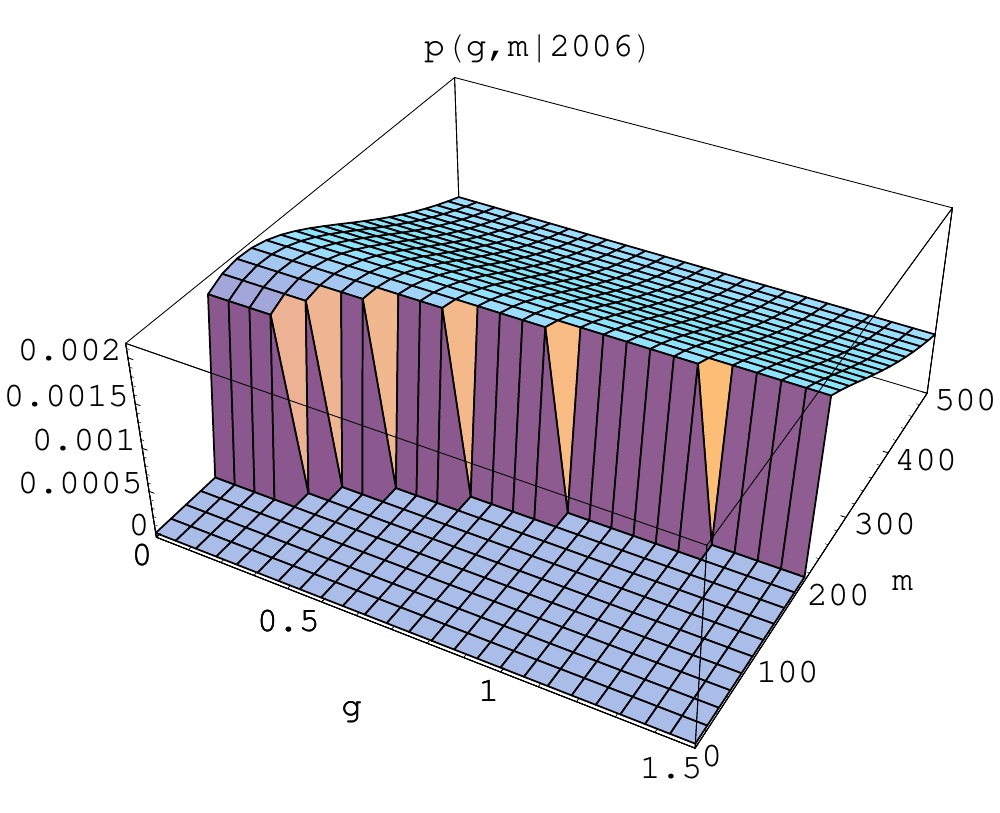} &
\includegraphics[width=2.1in]{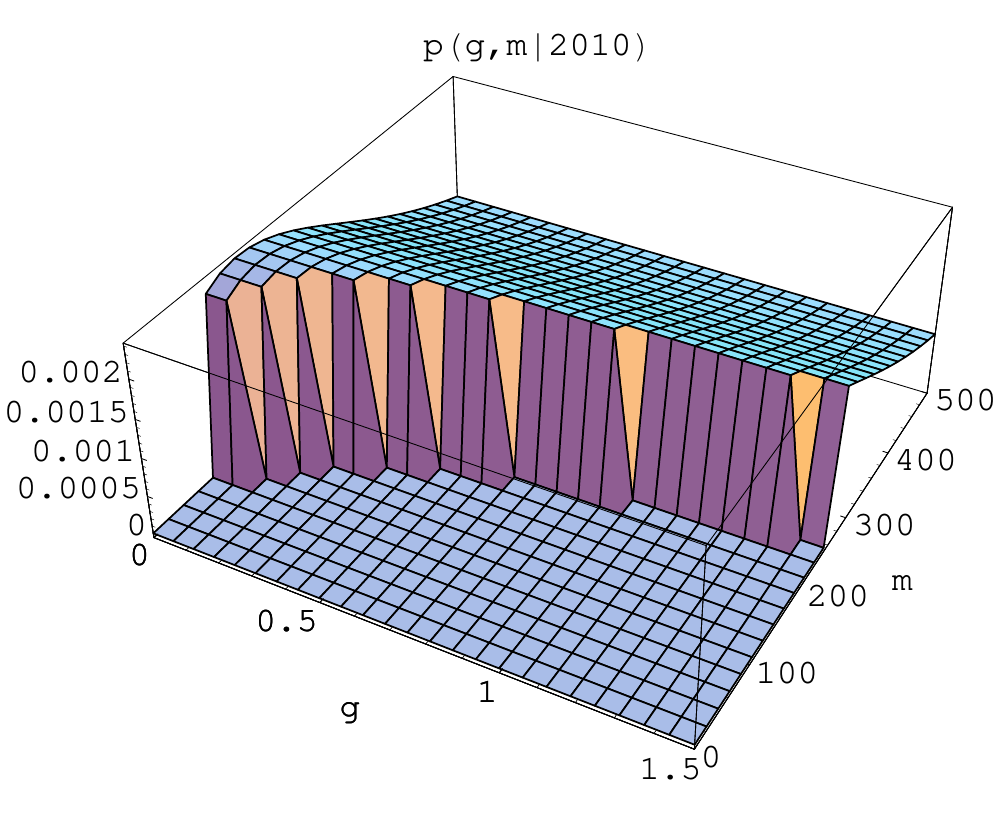} 
\end{tabular}
\caption{(a) The distribution $p(m|1990)$, representing 1990 expectation for the mass of the next non Standard Model particle to be discovered.  The scale of all mass axes is in units of GeV, and all distributions extend past the 500~GeV upper threshold used in constructing these plots.  (b) The distribution $p(g,m|1990)$, representing the joint prior expectation for the coupling $g$ and mass $m$ characterizing the next non Standard Model particle to be discovered.  (c) The distribution $p(g,m|1996)$, reflecting our prior from 1990 with the understanding gained from the null result of Tevatron Run I.  The effect of Tevatron Run I is to ``zero'' a substantial piece of the 1990 prior.  What is left is renormalized to unit integral.  This and subsequent plots show the case of pair production, with $n=2$.  The shape of the region zeroed by Tevatron Run I is determined considering the top quark as a representative benchmark for the sensitivity of Tevatron Run I in the variable space of coupling $g$ and mass $m$, as described in the text.  (d) The distribution $p(g,m|2000)$, reflecting the previous distribution with the additional knowledge of the null result from LEP\,2.  The effect of LEP\,2 is to zero that part of the density with $m<100$~GeV.  (e)  The distribution $p(g,m|2006)$, reflecting the previous distribution with the additional knowledge of the null result of Tevatron Run IIa.  The effect of Tevatron Run IIa is to zero $\approx 10\%$ of the density in the previous distribution. (f)  The distribution $p(g,m|2010)$, reflecting the previous distribution assuming a null result is obtained in Tevatron Run IIb.  The effect of Tevatron Run IIb is to zero $\approx 12\%$ of the density in the previous distribution.  All distributions (a)--(f) integrate to unity.  As a sanity check, note that the Standard Model Higgs ($g \approx 0.3$,$m \approx 130$) and single top production ($g \approx 0.3$,$m \approx 175$) are at the mode of $p(g,m|2006)$.}
\label{fig:pgmDistributions}
\end{figure*}

Prior probabilities can be defined given the field's knowledge in 1990, before Tevatron Run I (1990-1996) or LEP\,2 (1996-2000).  The production of a new process beyond the Standard Model is largely characterized by an overall coupling strength $g$ and the mass $m$ of the new state that is produced.  Our knowledge circa 1990 is thus characterized by a prior $p(g,m|1990)$ for the coupling $g$ and the mass $m$ of the most accessible non Standard Model physics process.  This prior is assumed to factorize, such that 
\begin{equation}
p(g,m|1990)=p(g|1990) \, p(m|1990).
\end{equation}
The prior for the coupling $p(g|1990)$ is taken to be uniform between 0 and 1.5, noting that for weak processes $g=\sqrt{4\pi\alpha_{\text{EM}}}\approx 0.3$ and that for strong processes $g=\sqrt{4\pi\alpha_{\text{s}}}\approx 1.2$.  Explicitly, 
\begin{equation}
p(g|1990)=(0<g<1.5)/1.5, 
\end{equation}
where $(0<g<1.5)$ evaluates to 1 if true, and 0 if false.
The prior $p(m|1990)$ for the mass $m$ of the most accessible new state is taken to be 
\begin{equation}
p(m|1990)= \frac{m}{\lambda^2} \exp{(-m/\lambda)},
\end{equation}
where $\lambda=150$~GeV sets the peak of the distribution $p(m|1990)$, as shown in Fig.~\ref{fig:pgmDistributions}(a).  This prior reflects the prejudice that new physics should appear at or slightly above the mass of the $W$ and $Z$ bosons.  The joint prior $p(g,m|1990)$ is shown in Fig.~\ref{fig:pgmDistributions}(b).

The proton and antiproton parton distrubtion functions are such that the single particle production cross section ($p\bar{p}\rightarrow X$) falls by a factor of two for every increase of $m$ by $\lambda_{1/2}=40$~GeV, holding $g$ constant.  By the same argument, the pair production cross section ($p\bar{p}\rightarrow X\bar{X}$) falls by a factor of two for every increase of $m$ by 20~GeV, holding $g$ constant.  Letting $n=1$ denote single particle production and $n=2$ denote pair production, the number of signal events $s$ varies with $g$, $m$, $n$, and the integrated luminosity $L$ according to 
\begin{equation}
\label{eqn:signalProductionCrossSection}
s \propto L \, g^2 \, 2^{-n m/\lambda_{1/2}}.
\end{equation}
The number of background events $b$ varies only with the integrated luminosity, so that $b \propto L$.  Combining these, the figure of merit is
\begin{equation}
\label{eqn:srootb}
\frac{s}{\sqrt{b}} \propto \sqrt{L} \, g^2 \, 2^{-n m/\lambda_{1/2}}.
\end{equation}

In Tevatron Run I, the top quark was discovered at the edge of experimental sensitivity.  This discovery corresponded to the strong coupling $g=1.2$, a mass $m$ equal to the top quark mass $m_{t}=175$~GeV, pair production of $n=2$, and an integrated luminosity of $L \approx 100$~pb$^{-1}$.  Inserting these numbers into Eq.~\ref{eqn:signalProductionCrossSection} yields 
\begin{equation}
\label{eqn:srootb_top}
\left.\frac{s}{\sqrt{b}}\right|_{\text{top}} \propto \sqrt{100~\text{pb}^{-1}} \, (1.2)^2 \, 2^{-2 m_{t}/\lambda_{1/2}}.
\end{equation}
Taking the top quark discovery as a relevant benchmark, a new process characterized by $g$, $m$, and $n$ will be observed at the Tevatron after collecting an integrated luminosity $L$ if
\begin{equation}
\label{eqn:conditionForDiscoveringANewProcessAtTheTevatron}
\sqrt{\frac{L}{100~\text{pb}^{-1}}} \left(\frac{g}{1.2}\right)^2 \frac{2^{-n m/\lambda_{1/2}}}{2^{-2 m_{t}/\lambda_{1/2}}} >1,
\end{equation}
where this expression is obtained by dividing Eq.~\ref{eqn:srootb} by Eq.~\ref{eqn:srootb_top}.
Equation~\ref{eqn:conditionForDiscoveringANewProcessAtTheTevatron} implies that given the existence of a new physics process characterized by a coupling $g$, a mass scale $m$, and single or pair production $n$, the probability of observing a null result at the Tevatron upon the accumulation of integrated luminosity equal to $L$ is
\begin{equation}
\label{eqn:tevatronLikelihood}
p({\text{tev}}(L)|1990,g,m) 
 = \left(g<\frac{1.2\times 2^{(nm/2-m_{t})/\lambda_{1/2}}}
                {\left(\frac{L}
                            {100~\text{pb}^{-1}}
                 \right)^{1/4}}
   \right),
\end{equation}
where the right hand side is just a rewriting of Eq.~\ref{eqn:conditionForDiscoveringANewProcessAtTheTevatron}, using the notation that the boolean expression evaluates to 1 if true, and 0 if false.  

The expectation in 1990 that new physics would be seen at Tevatron Run I or LEP\,2 is determined to be
\begin{equation}
1-p(\text{tev1},\text{lep2}|1990) =
30\%.
\end{equation}

Letting $p(g,m|1996)$ denote our knowledge in 1996, comprising the prior from 1990 plus the additional knowledge (${\text{tev1}}={\text{tev}}(100~\text{pb}^{-1})$) of no processes beyond the Standard Model having been observed in Tevatron Run I, from application of Bayes's theorem
\begin{eqnarray}
\label{eqn:pgm1996}
p(g,m|1996) & = & p(g,m|1990,{\text{tev1}}) \nonumber \\
            & = & \frac{p(g,m,{\text{tev1}}|1990)}{p({\text{tev1}}|1990)} \nonumber \\
            & = & \frac{p({\text{tev1}}|1990,g,m) p(g,m|1990)}{p({\text{tev1}}|1990)}.
\end{eqnarray}
Here $p({\text{tev1}}|1990,g,m)$ is defined by Eq.~\ref{eqn:tevatronLikelihood}, and $p({\text{tev1}}|1990)$ quantifies the probability of having obtained the null Tevatron Run I result given our knowledge in 1990, representing the density left in going from Fig.~\ref{fig:pgmDistributions}(b) to Fig.~\ref{fig:pgmDistributions}(c), which shows a plot of $p(g,m|1996)$.

Summarizing the null result from LEP\,2 as ruling out the possibility that $m/n<200$~GeV leads to 
\begin{equation}
p({\text{lep2}}|1996,g,m) = (\frac{m}{n}>200~\text{GeV}),
\end{equation}
using again the notation that the boolean expression evaluates to 1 if true, and 0 if false.  Letting $p(g,m|2000)$ denote our understanding in 1996 with the additional knowledge ({\text{lep2}}) of no processes beyond the Standard Model having been observed in LEP\,2, a similar application of Bayes's theorem yields
\begin{equation}
p(g,m|2000) = \frac{p({\text{lep2}}|1996,g,m) p(g,m|1996)}{p({\text{lep2}}|1996)}.
\end{equation}
Here $p(g,m|1996)$ is taken from Eq.~\ref{eqn:pgm1996}, and $p({\text{lep2}}|1996)$ represents the density left in going from Fig.~\ref{fig:pgmDistributions}(c) to Fig.~\ref{fig:pgmDistributions}(d).  The latter shows a plot of $p(g,m|2000)$, showing that the region with $m<100$~GeV has been zeroed.

Letting $p(g,m|2006)$ denote our understanding in 2006, comprising our knowledge in 2000 plus the additional knowledge ($\text{tev2a}=\text{tev}(1~\text{fb}^{-1})$) of no signal beyond the Standard Model having been observed in Tevatron Run IIa, the posterior distribution $p(g,m|2006)$ obtained is shown in Fig.~\ref{fig:pgmDistributions}(e).  The fraction of the density left in going from Fig.~\ref{fig:pgmDistributions}(d) to Fig.~\ref{fig:pgmDistributions}(e) is $p(\text{tev2a}|2000) \approx 90\%$.

Letting $p(g,m|2010)$ denote our understanding in 2010, comprising our knowledge in 2006 and assuming ($\text{tev2b}=\text{tev}(10~\text{fb}^{-1})$) that no processes beyond the Standard Model are observed in Tevatron Run IIb, the posterior distribution $p(g,m|2010)$ obtained is shown in Fig.~\ref{fig:pgmDistributions}(f).  The fraction of the density left in going from Fig.~\ref{fig:pgmDistributions}(e) to Fig.~\ref{fig:pgmDistributions}(f) is $p(\text{tev2b}|2006) \approx 88\%$.

From the calculations above, in 2000 the expectation for finding new physics in Tevatron Run IIa was equal to 
\begin{equation}
(1-p({\text{tev2a}}|2000)) \approx 10\%.
\end{equation}
In 2006, the expectation for finding new physics in Tevatron Run IIb is 
\begin{equation}
(1-p({\text{tev2b}}|2006)) \approx 12\%.
\end{equation}  
In 2000, the expectation for finding new physics in Tevatron Run II is equal to 
\begin{eqnarray}
(1-p({\text{tev2a}},{\text{tev2b}}|2000)) &=& \nonumber \\
(1-p({\text{tev2a}}|2000) \,p({\text{tev2b}}|2006)) &\approx& 20\%.
\end{eqnarray}
Carrying the argument forward, the expectation for finding new physics at the LHC is determined to be roughly 90\%.

To recap, the physics content of the argument provided above is straightforward.  Starting with a prior representing the field's expectation in 1990 that the first sign of physics beyond the Standard Model will appear at or slightly above the electroweak scale, this prior is evolved through the experimental knowledge gained from Tevatron Run I and LEP\,2 to update our expectation to the year 2000.  This expectation is then evolved through the experimental reality of Tevatron Run IIa and a hypothetical null result in Tevatron Run IIb to find that the fraction of our expectation from 2000 that has been eliminated by 2010 is $\sim 20\%$. 

Of the roughly $2 \times 10^4$ kinematic distributions that can be considered in the Tevatron, it is difficult to convincingly argue that any one of them is more likely to display the first sign of new physics than any other.  Unless there are many signs of new physics in the Tevatron data, it therefore follows that any targeted search --- including the search for single top production, $B_s\rightarrow \mu^+\mu^-$, and the measurement of $B_s$ mixing --- has a discovery expectation of roughly $20\%/2 \times 10^4 = 10^{-5}$.

\subsection{$\mu2e$}
\label{sec:DiscoveryExpectationAppendix:mu2e}

The argument for the $\mu2e$ experiment is largely based on the observation that new electroweak scale physics with non-negligible lepton flavor violation can give rise to a transition $\mu\rightarrow e$ in the electric field of an atomic nucleus sufficiently large to be observable at $\mu2e$.  Of the new electroweak scale physics that may give rise to such a signal, electroweak scale supersymmetry is so far the most thoroughly studied possibility.

A poll taken during a series of seminars over the past six years including over six hundred professional physicists~\footnote{The seminar audience is asked whether the first sign of new physics will come from heavy gauge bosons, technicolor, extra spatial dimensions, leptoquarks, supersymmetry, a fourth generation of fermions, compositeness, or something else.  The options are provided on a transparency, and audience members are requested to raise their hand as the above possibilities are read in sequence.  The fraction of audience members who raise their hand for something, corresponding to a response rate for this poll, is roughly 50\%.} suggests the field's expectation that weak scale supersymmetry is realized in nature is roughly 35\%.  An independent poll conducted online shows similar results~\cite{ChoudalakisPoll}.  Interestingly, it is generally not the case that one third of the field is confident low energy supersymmetry is present, and that two thirds are confident low energy supersymmetry is not; informal discussions suggest most individual physicists would place the odds that the first sign of new physics will arise from low energy supersymmetry at 2:1 against.

Assuming low energy supersymmetry is realized, either there is significant mixing in the slepton sector (as in the case of neutrinos in the Standard Model), or for some reason this mixing is suppressed (as in the case of quarks in the Standard Model).  There being no particularly compelling reason to prefer either of these two scenarios over the other, an expectation of 50\% can be assigned to there being significant mixing in the slepton sector, asssuming low energy supersymmetry is realized in nature.

Assuming $\mu2e$ operates as designed, the expectation that a $\mu\rightarrow e$ signal identifying physics beyond the Standard Model will be seen is thus roughly $35\% \times 50\% = 18\%$.  

Reasonable variations in this argument, including expanding the consideration of new physics scenarios that may lead to a $\mu\rightarrow e$ signal beyond electroweak scale supersymmetry, can lead to a final number differing by roughly a factor of two either way.  The important point is that the discovery expectation cannot reasonably be argued to be significantly less than $\sim 10\%$ or greater than $\sim 90\%$~\footnote{The author's initial assessment of the probability of a signal being observed at $\mu2e$ was $\sim10^{-3}$, for reasons that did not withstand interrogation.  This is an example of a case in which an individual's assessment of the scientific merit of an experiment has changed qualitatively through forced exposition of quantitative argument.}.

For a proposed experiment whose outcome is binary, a prior expectation of roughly 50\% corresponds to a scientific merit of $0.30$, and the experiment provides a full bit of information.  A proposed experiment with binary outcome with a prior expectation to see a signal of 10\% (or 90\%) has a scientific merit of $0.14$.  The scientific merit of the proposed $\mu2e$ experiment is significantly less than the LHC experiments in part due to the relative inability of the $\mu2e$ experiment to further explore the underlying physics responsible for producing the signal.

\subsection{Historical discoveries}

The surprises corresponding to the discoveries of the $c$, $\tau$, $b$, $W$, $Z$, and $t$ particles have been estimated by communication with individuals active (both intimately and as observers) during the time of the discoveries.  Written documents that have been helpful in this regard include Refs.~\cite{Pickering,Feldman:1992vk,Taubes}.

\bibliography{scientificMerit}

\end{document}